\documentclass[prd,reprint,amsmath,amssymb,aps]{revtex4-2}
\input ./belle2sym.tex
\usepackage{graphicx}
\usepackage{dcolumn}
\usepackage{bm}
\usepackage{hyperref}
\usepackage{orcidlink}
\usepackage{relsize}
\usepackage{tikz}
\usepackage{graphicx}
\usepackage{overpic}  
\usetikzlibrary{positioning}

\def\babar{\mbox{\slshape B\kern-0.1em{\smaller A}\kern-0.1em B\kern-0.1em{\smaller A\kern-0.2em R}}\xspace}

\newcommand{\btodlplm}{\ensuremath{b\to d\, \ell^+\ell^-}\xspace}

\newcommand{\btoslplm}{\ensuremath{b\to s\, \ell^+\ell^-}\xspace}

\newcommand{\ctoulplm}{\ensuremath{c\to u\, \ell^+\ell^-}\xspace}

\newcommand{\lplm}{\ensuremath{l^+l^-}\xspace}

\newcommand{\hhee}{\ensuremath{h^- h^{(\prime)+}e^+e^-}\xspace}

\newcommand{\mhhee}{\ensuremath{m_{h h^{(\prime)}ee}}\xspace}
\newcommand{\hheered}{\ensuremath{h h^{(\prime)}ee}\xspace}

\newcommand{\kpiee}{\ensuremath{K^-\pi^+e^+e^-}\xspace}

\newcommand{\kkee}{\ensuremath{K^+K^-e^+e^-}\xspace}

\newcommand{\pipiee}{\ensuremath{\pi^+\pi^-e^+e^-}\xspace}

\newcommand{\kthreepi}{\ensuremath{K^-\pi^+\pi^-\pi^+}\xspace}

\newcommand{\kkpipi}{\ensuremath{K^+K^-\pi^+\pi^-}\xspace}

\newcommand{\mee}{\ensuremath{m_{ee}}\xspace}

\newcommand{\deltam}{\ensuremath{\Delta m}\xspace}
\newcommand{\Dscmp}{\ensuremath{p^*(D^{*+})}\xspace}

\begin{document}
\title{Measurement of the \Dz\ra\kpiee branching fraction and search for \Dz\ra\pipiee and \Dz\ra\kkee decays at Belle}
  \author{I.~Adachi\,\orcidlink{0000-0003-2287-0173}} 
  \author{L.~Aggarwal\,\orcidlink{0000-0002-0909-7537}} 
  \author{H.~Ahmed\,\orcidlink{0000-0003-3976-7498}} 
  \author{Y.~Ahn\,\orcidlink{0000-0001-6820-0576}} 
  \author{H.~Aihara\,\orcidlink{0000-0002-1907-5964}} 
  \author{N.~Akopov\,\orcidlink{0000-0002-4425-2096}} 
  \author{S.~Alghamdi\,\orcidlink{0000-0001-7609-112X}} 
  \author{M.~Alhakami\,\orcidlink{0000-0002-2234-8628}} 
  \author{A.~Aloisio\,\orcidlink{0000-0002-3883-6693}} 
  \author{N.~Althubiti\,\orcidlink{0000-0003-1513-0409}} 
% ...
\collaboration{The Belle and Belle II Collaborations}

\begin{abstract}
We present a study of the rare charm meson decays $\Dz\ra\kkee$, $\pipiee$, and $\kpiee$ using a 942 \invfb data set collected by the Belle detector at the KEKB asymmetric-energy \epem collider. We identify \Dz candidates via the charge of the pion from $\Dstarp\ra\Dz\pip$ decays and normalize the branching fractions to \Dz\ra\kthreepi decays. The branching fraction for decay \Dz\ra\kpiee is measured to be (39.6 $\pm$ 4.5 (stat) $\pm$ 2.9 (syst)) $\times$ $10^{-7}$, with the dielectron mass in the $\rho/\omega$ mass region $675 < \mee < 875 \mevcc$. 
We also search for \Dz\ra\hhee ($h^{(\prime)}=K,\,\pi$) decays with the dielectron mass near the $\eta$ and $\phi$ resonances, and away from these resonances for the \kkee and \pipiee modes. For these modes, we find no significant signals and set 90\% confidence level upper limits on their branching fractions at the $\mathcal{O}$(10$^{-7}$) level. 
\end{abstract}
\maketitle
\begin{figure}
   \begin{overpic}[width=\linewidth]{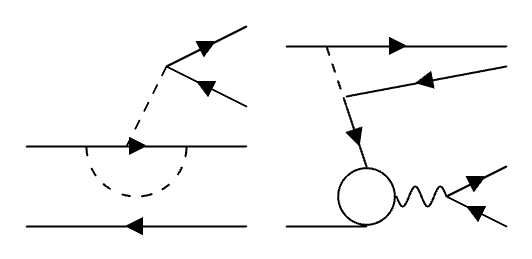}
   \put(2,5){\text{\ubar}} 
   \put(47.5,5){\text{\ubar}} 
   \put(2.1,20.5){\text{$c$}}
   \put(23,14){\text{$W^+$}}
   \put(18.5,28){\text{$\g/\Z$}}
   \put(47.5,27.3){\text{$l^+$}}
   \put(47.5,42.5){\text{$l^-$}}
   \put(47.5,20.5){\text{$u$}}
   \put(51.2,5.){\text{$\ubar$}} 
   \put(51.2,39.){\text{$c$}} 
   \put(96.7,39.){\text{$d$}}
   \put(96.7,34.5){\text{$\dbar$}}
   \put(55.2,33.){\text{$W^+$}}
   \put(62.5,20.5){\text{$u$}}
   \put(68,10.8){\text{$\rho$}}
   \put(78.5,15){\text{$\g$}}
   \put(78.5,15){\text{$\g$}}
   \put(96.7,16.3){\text{$l^-$}}
   \put(96.7,5){\text{$l^+$}}
    \end{overpic}
\caption{SD contributions to FCNC \Dz decays through an electroweak penguin diagram (left). LD contributions to $\Dz\ra X^0V^0\ra X^0 \lplm$ decays through the VMD diagram (right). \label{fig-SMdiagrams-SD}}
\end{figure}

Electroweak penguin quark transitions mediated by flavor changing neutral currents (\mbox{FCNCs}) such as \btoslplm, \btodlplm, and \ctoulplm (where $\ell^\pm$ is an electron or muon) are forbidden at tree level in the standard model (SM) \footnote{The inclusion of the charge-conjugate decay mode is implemented throughout the letter unless otherwise stated.}. 
Within the SM, FCNCs proceed via loop diagrams and therefore are highly suppressed. Thus, \ctoulplm decays probe beyond the standard model (BSM) physics that could affect the decay rate and other variables. The BSM amplitudes can interfere with the SM amplitudes, altering physics observables from the SM predictions such as total and differential decay rates, and affecting tests of lepton flavor universality~\cite{PhysRevD.98.035041,PhysRevD.83.114006,PhysRevD.76.074010,Cappiello2013,Bause2020}.\\
\indent The decays \ctoulplm are FCNC transitions of a charm quark to an up quark and a lepton pair. Compared to \btoslplm and \btodlplm decays, these transitions are further suppressed due to the Glashow-Iliopoulos-Maiani mechanism and the small quark masses relative to the top quark in the loop~\cite{Burdman:2001tf}. 
The decays $D^0\rightarrow X^0 \ell^+\ell^-$, where $X^0$ is a light-quark system, can have contributions from both 
short-distance (SD) and long-distance (LD) amplitudes, as shown in Fig.~\ref{fig-SMdiagrams-SD}. The SD decay amplitudes are suppressed, with branching fractions ($\mathcal{B}$) reaching only the $2 \times 10^{-9}$ level~\cite{Fajfer:2002bu}. 
However, LD contributions from photon pole or vector meson dominance (VMD) amplitudes, which proceed through the decays 
$D^0 \rightarrow X^0 (\gamma^*/V^0)\rightarrow X^0 \ell^+\ell^-$, where $\gamma^*$ is an off-shell virtual photon and $V^0$ is an intermediate vector meson ($\rho,\,\omega,\,\phi$), can reach values of up to 2 $\times$ 10$^{-6}$~\cite{Fajfer:2002bu} for the Cabibbo-favored decay \Dz\ra\kpiee.
 \\
\indent Several BSM scenarios such as the minimal supersymmetric standard model, models including leptoquarks, little Higgs, $Z'$ models, and models with warped extra dimensions predict significantly enhanced rates for \ctoulplm decays~\cite{PhysRevD.98.035041,PhysRevD.83.114006,PhysRevD.76.074010,Cappiello2013,Bause2020,PhysRevD.109.036027, Bause:2019vpr, deBoer:2015boa}. Thus, measurements of branching fractions for these decays allow us to probe for BSM physics and to characterize the LD contributions to the decay amplitudes.\\
\indent The \babar~\cite{Lees:2011hb,Lees:2018vns,Lees:2019pej,Lees:2020qgv}, BES~III~\cite{Ablikim:2018gro}, CLEO~II~\cite{Freyberger:1996it}, D0~\cite{PhysRevLett.100.101801},
Fermilab E653~\cite{Kodama:1995ia}, E791~\cite{Aitala:2000kk}, and LHCb~\cite{Aaij:2015hva,Aaij:2017iyr,Aaij:2017nsd,JHEP06(2021)044,LHCb2025} Collaborations 
have searched for rare and forbidden $X_c \to h\, (h^{(\prime)})\, \ell^+ \ell^{-(\prime)}$ decays in several final states. BES~III sets upper limits (UL) at the 90\% confidence level (CL) in the range $(11 - 41)$ $\times$ 10$^{-6}$ for $D^{0} \rightarrow$ \hhee decays~\cite{Ablikim:2018gro}. Recently, several four-body decays $D^0 \to h h^{(\prime)} \ell^+ \ell^-$ (where 
$hh^{(\prime)} = KK, \pi\pi, K\pi$) have been observed.  \babar observed the decay $D^0\rightarrow K^-\pi^+e^+e^-$ in the mass range 675 $<$ \mee $<$ 875 \mevcc at a rate compatible with VMD contributions, and set a branching fraction upper limit on $\Dz\ra \kpiee$, excluding $\epem$ resonances with branching fractions above $3.1\times 10^{-6}$ at the 90\% confidence level~\cite{Lees:2018vns}. LHCb observed the decay $D^0\rightarrow K^-\pi^+\mu^+\mu^-$~\cite{Aaij:2015hva}, and also observed the decays $D^0\rightarrow \pi^+\pi^-\mu^+\mu^-$ and $D^0\rightarrow K^+K^-\mu^+\mu^-$~\cite{Aaij:2017iyr}.\\
\indent Here we search for the rare charm meson decays $\Dz\ra\kkee$, $\pipiee$, and $\kpiee$ using data collected by the Belle experiment. We analyze the \epem \ra \ccbar data that has a total integrated luminosity of 942 \invfb. The data was collected at center-of-mass energies ($E_{\text{cm}}$) at the \FourS resonances or 60 \mev below, at the \FiveS resonance, and in the $10860< E_{\text{cm}} < 11020$ \mev energy scan. The data was recorded from 2000 to 2010 from the collision of 8~GeV electrons
with 3.5~GeV positrons at the KEKB collider \cite{AKAI2003191}. The
Belle detector, a large-solid-angle magnetic spectrometer,
is described in detail elsewhere \cite{KUROKAWA20031}. The Belle inner detector consists of a four-layer silicon vertex detector, a 50-layer central
drift chamber, an array of aerogel threshold
Cherenkov counters, a barrel-like arrangement of
time-of-flight scintillation counters, and an electromagnetic
calorimeter composed of CsI (Tl) crystals,
all located inside a superconducting solenoid coil that
provides a 1.5 T magnetic field. An iron flux-return yoke placed outside the coil is instrumented with resistive plate chambers to detect $\KL$ mesons and muons. \\
\indent We use Monte Carlo (MC) simulated events to optimize selection criteria, calculate reconstruction efficiencies, and study background sources. We generate the MC event samples using {\tt{EvtGen}} \cite{Lange:2001uf}, {\tt{PYTHIA}} \cite{Sjostrand:2003wg}, and we use {\tt{PHOTOS}} \cite{Photos} and {\tt{Geant3}} \cite{Brun:geant3} to simulate final state radiation and the detector response, respectively. For each signal channel we generate $hh^{(\prime)}$ and $ee$ resonant and non-resonant signal MC samples. We neglect interference between non-resonant and resonant decays. We use MC samples of \epem\ra\qqbar (where $q = u, d, s$ or $c$) and \epem\ra\BB corresponding to six times that of the data to study the background composition.\\
\indent We require at least five charged tracks in the event. Each track must have a momentum greater than 0.1\gevc. We require the distance of the closest approach to the origin to be less than 4.5 cm along the beam direction and less than 0.25 cm transverse to the beam direction to reduce beam-induced backgrounds and background from \KS mesons. We perform particle identification (PID) based on information provided by detector subsystems in the form of likelihoods $\mathcal{L}_{i}$ for species $i$, where $i = e,\,\mu,\,\pi,\, K,$ or \proton for each track. Kaon candidates must have ${\mathcal R}_{K} = \mathcal{L}_{K}/(\mathcal{L}_{K} + \mathcal{L}_{\pi})$ $>$ 0.1 for the \kpiee and \kkee mode, and pion candidates are required to have ${\mathcal R}_{K}<$ 0.4 for the \pipiee mode. These requirements have kaon and pion identification efficiencies of about 97\% and 91\%, with misidentification rates of about 20\% and 10\%, respectively. The electron candidates must have ${\mathcal R}_{e}$ = $\mathcal{L}_{e}/(\mathcal{L}_{e} + \mathcal{L}_{\mu} + \mathcal{L}_{K} + \mathcal{L}_{\pi} + \mathcal{L}_{\proton})$ $>$ 0.8. To recover electron bremsstrahlung, we add photon(s) having a minimum energy of 20\mev and an angle within five degrees around the direction of the electron track at the interaction point to the four-momenta of the electron candidate. The electron identification efficiency is about 91\%, with a misidentification rate of less than 3\%. We use the B2BII software package~\cite{Gelb:2018agf} to convert the Belle data to Belle II data format and analyze the data with the Belle II analysis software framework ({\tt basf2})~\cite{Kuhr:2018lps}.

We reconstruct $\Dz\ra\kpiee$, $\pipiee$, and $\kkee$ signal candidates from the selected kaon, pion, and electron candidates. Candidates with \Dz invariant mass \mhhee in the range 1.80 $< \mhhee <$ 1.93 \gevcc are combined with a $\pip$ candidate to form a \Dstarp candidate. The requirement of a \Dstarp tagged \Dz suppresses the background from random track combinations. Candidates must have a \Dstarp momentum in the center-of-mass frame $\Dscmp > 2.5$ \gevc to reduce the combinatorial background from \B decays and a mass difference between \Dstarp and \Dz candidates $\deltam$ within 0.5 \mevcc of the nominal value~\cite{Workman:2022ynf} to be consistent with the decay $\Dstarp\to\Dz \pi^+$. We also apply a vertex fit to the decay chain $\Dstarp \ra \Dz \pi^+$, \Dz\ra\hheered, with the \Dstar production vertex constrained to the interaction point. We discard candidates that fail this fit. 

The decay $\piz/\eta\ra\epem\gamma$ can produce a complicated background shape, which is difficult to model. The electron bremsstrahlung recovery can mistakenly include a photon originating from a $\piz/\eta$ decay so that the reconstructed $m(\hheered\gamma)$ will fake the signal. Such decays will also contribute to a background in the \mhhee region below the \Dz mass, resulting in a nonlinear background shape. To suppress these backgrounds, we apply the selection $\mee > 200$ \mevcc. In addition, we do not apply electron bremsstrahlung recovery for candidates with \mee in the $\eta$ mass region (520, 560\mevcc). 

Some candidates include electrons originating from photon conversions. In order to veto these events, we combine the $e^\pm$ from the signal (\Dz) candidate with another oppositely charged track from the event to form a candidate \epem pair. We require a converged vertex fit for the photon conversion candidates $(\epem)$ and discard the corresponding \Dz candidate if the angle between the $\epem$ tracks in the lab frame is less than 0.07 radians or the invariant mass of $\epem$ tracks is less than 100 \mevcc. 

Hadronic \Dz decays in which one or more of the \Dz daughters is misidentified as a lepton also contribute to the background. In each event, we reconstruct $\Dstarp\to\Dz \pi^+_s$ with $\Dz \ra \kthreepi$, $\Dz \ra \pi^+\pi^-\pi^+\pi^-$, and $\Dz \ra \kkpipi$ decays in addition to the signal modes \ensuremath{h h^{(\prime)}e^+e^-}\xspace. We discard the corresponding signal candidate if any of the reconstructed hadronic \Dz decay candidates have invariant mass and \deltam within 3 \mevcc and 0.4 \mevcc, respectively, of the corresponding nominal values.

For each signal mode, we optimize the selection criteria for \Dscmp, \deltam, PID, photon conversion and hadronic \Dz vetos in the $\eta$ (520, 560), $\rho/\omega$ (675, 875), and $\phi$ (990, 1035\mevcc) mass regions in order to search for potential $ee$ resonant decays. We also search for \Dz\ra\hhee decays in the \mee spectrum not included in the resonant regions defined above which we refer to as "non-resonant". The \mee regions mentioned above are not individually optimized, with the \mee ranges covering about 80\% of the corresponding $ee$ resonance regions. For events with more than one signal candidate, the candidate with $\deltam$ closest to the known value is selected. We optimize the cuts by maximizing a figure-of-merit ${S}/{\sqrt{S+B}}$ for each \mee region, where $S$ and $B$ are the expected number of signal candidates in data estimated from PDG~\cite{Workman:2022ynf} branching fractions and background yields estimated using background MC samples, respectively. Since the \Dz production rate is not precisely known, we measure the signal branching fractions relative to the normalization decay \Dz\ra\kthreepi, with similar selections applied such as PID. 

We calculate the signal branching fractions and upper limits using the equation
\begin{equation}
\mathcal{B}(\hheered)=\frac{N_{\hheered}}{N_{K\pi\pi\pi}}\frac{\epsilon_{K\pi\pi\pi}}{\epsilon_{\hheered}}\mathcal{B}(\kthreepi),
\end{equation}
where $N$ are the yields, and $\epsilon$ are the reconstruction efficiencies. We measure the branching fractions or set branching fraction upper limits for various \mee regions in each \hheered mode.

We use a one-dimensional unbinned extended maximum likelihood fit to \mhhee to extract the signal yield for each decay mode in the $\eta$ (520, 560), $\rho/\omega$ (675, 875), $\phi$ (990, 1035\mevcc), and remaining \mee regions. The signal probability density function (PDF) is a Gaussian-like function with different resolutions above and below the \Dz mass \footnote{The normalization mode signal PDF is a sum of the signal PDF used for the signal channels and a Gaussian with a shared mean for the \Dz mass.}. We obtain the signal PDF parameters from fits to the signal MC distributions, and we fix these parameters for the signal yield extraction. We model the background using a linear function, where the slope parameter is floated in the fit. We do not examine any signal mode distributions until the analysis procedure is finalized to minimize potential biases on the measured quantities.

\par
\begin{figure}[!h]
\begin{tikzpicture}
\hspace{-0.36cm}
  \draw[=base,color =white] (155,0) -- (155,0)
   node[color = black,rotate=90,shift={(-0.,0.)}] at (120,0) {\scriptsize{{Events $/ (5 \mevcc)$}}};
\end{tikzpicture}
  \begin{minipage}{0.24\textwidth}
  \vspace{-2.5cm}
    \includegraphics[width=\textwidth]{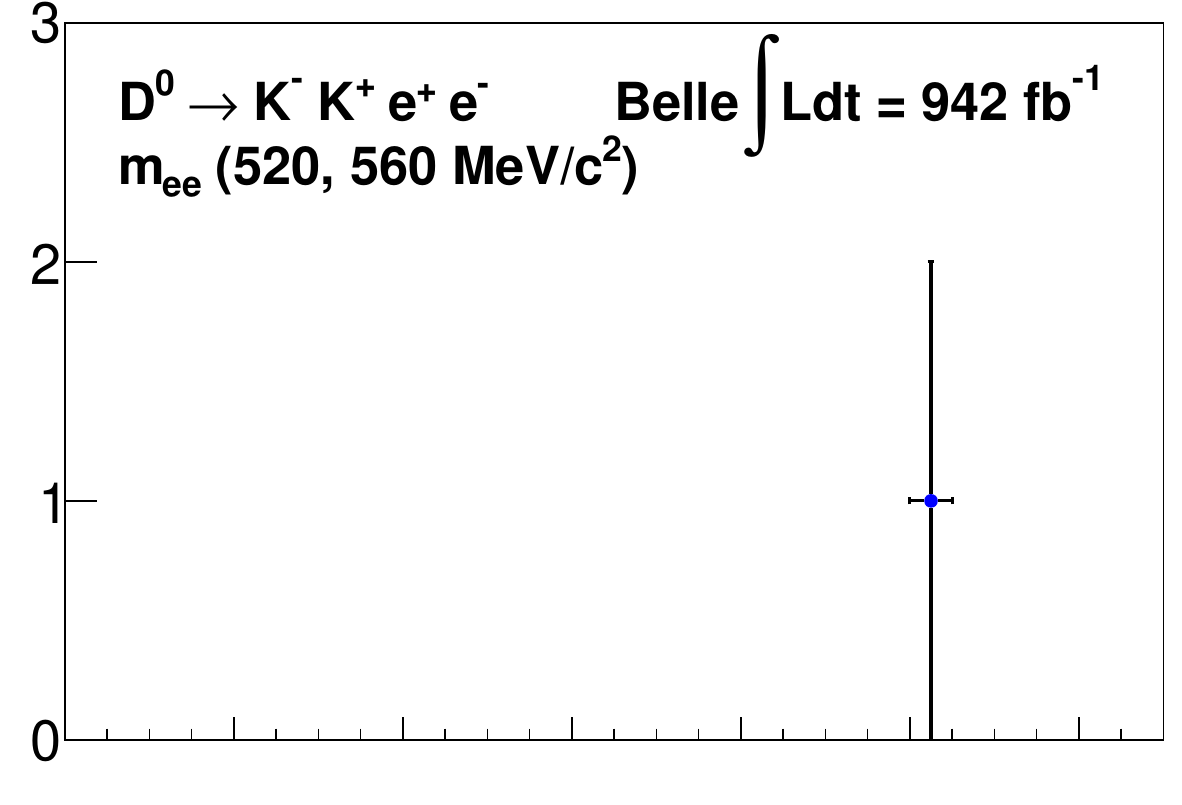} 
  \end{minipage}
\hspace{-0.26cm} %
  \begin{minipage}{0.24\textwidth}
  \vspace{-2.5cm}
    \includegraphics[width=\textwidth]{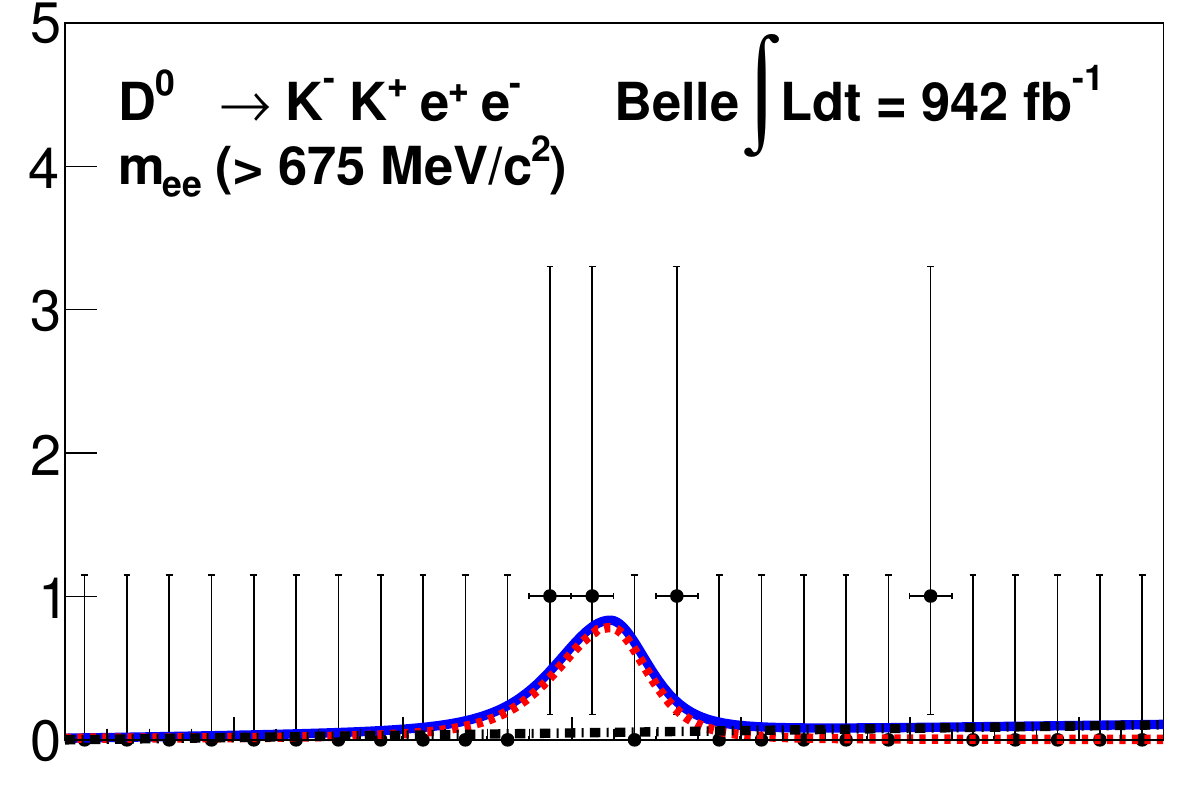}
  \end{minipage}
  \begin{minipage}{0.24\textwidth}
  \vspace{-0.19cm}
    \includegraphics[width=\textwidth]{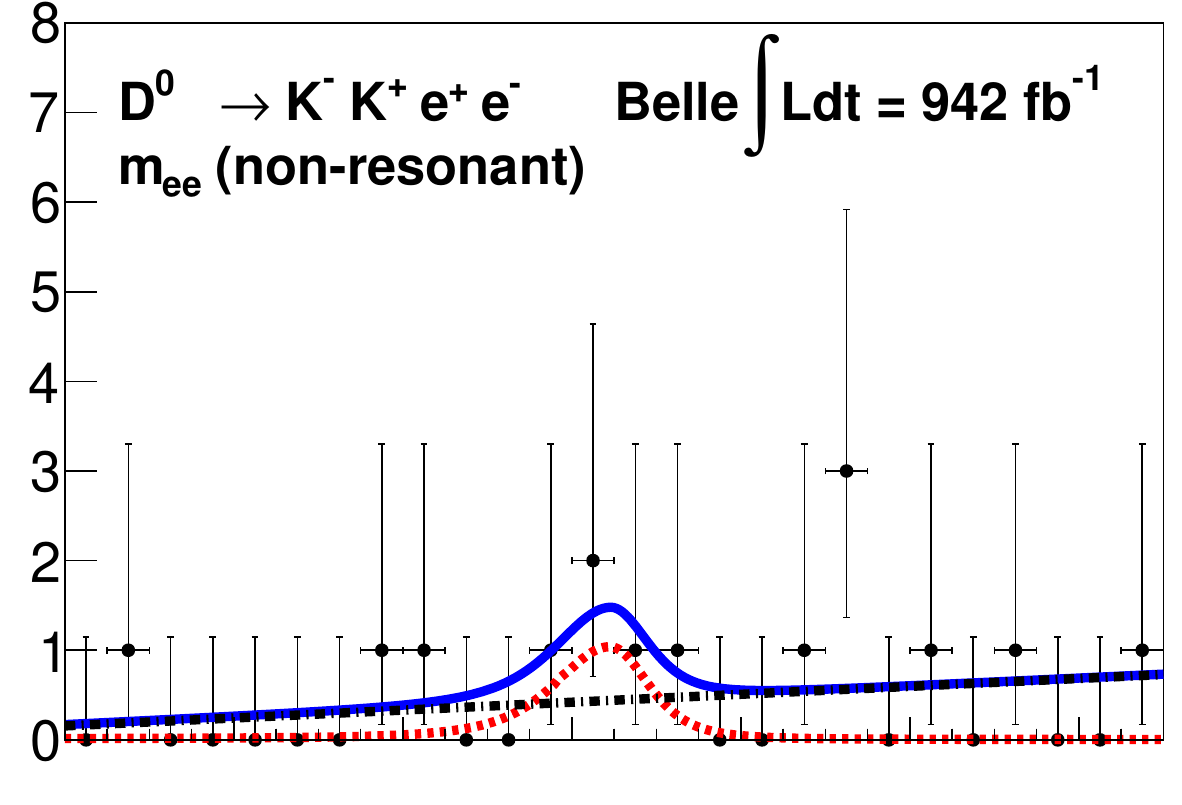}
  \end{minipage}
\hspace{-0.26cm} %
  \begin{minipage}{0.24\textwidth}
  \vspace{-0.19cm}
    \includegraphics[width=\textwidth]{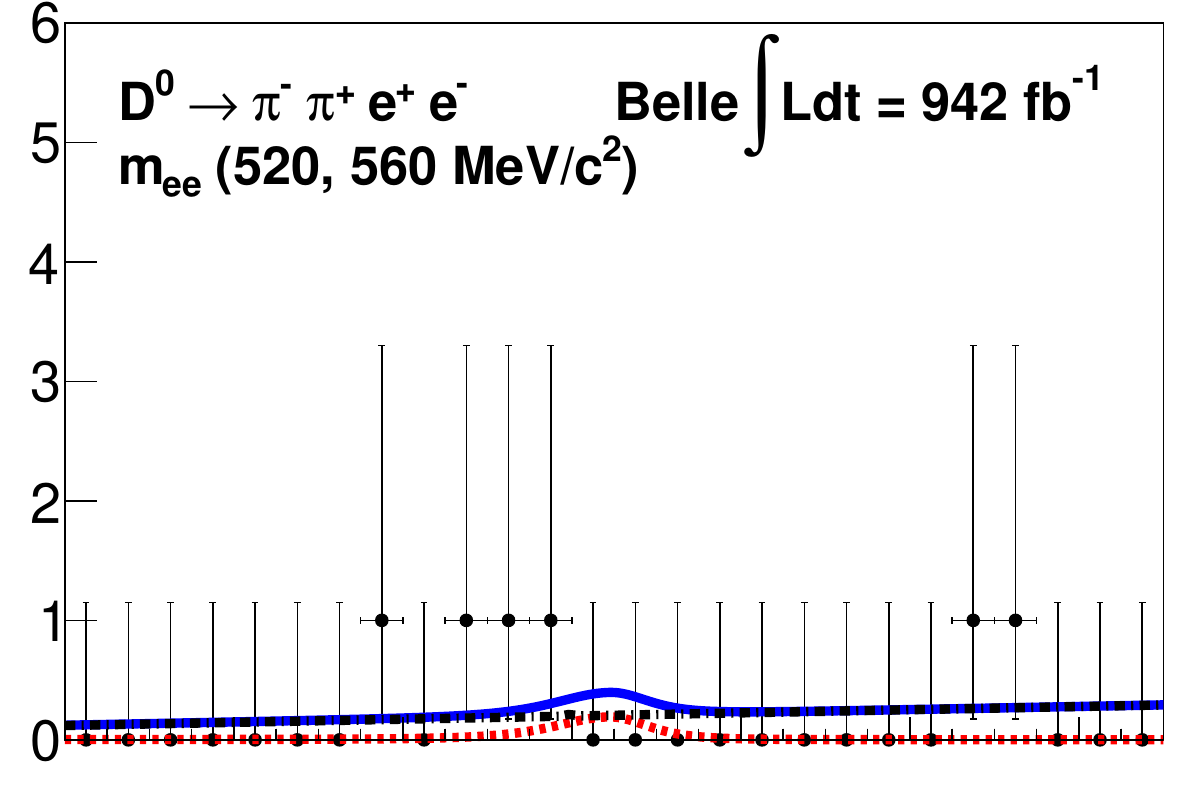}
      \end{minipage}
    \begin{minipage}{0.24\textwidth}
    \vspace{-0.19cm}
    \includegraphics[width=\textwidth]{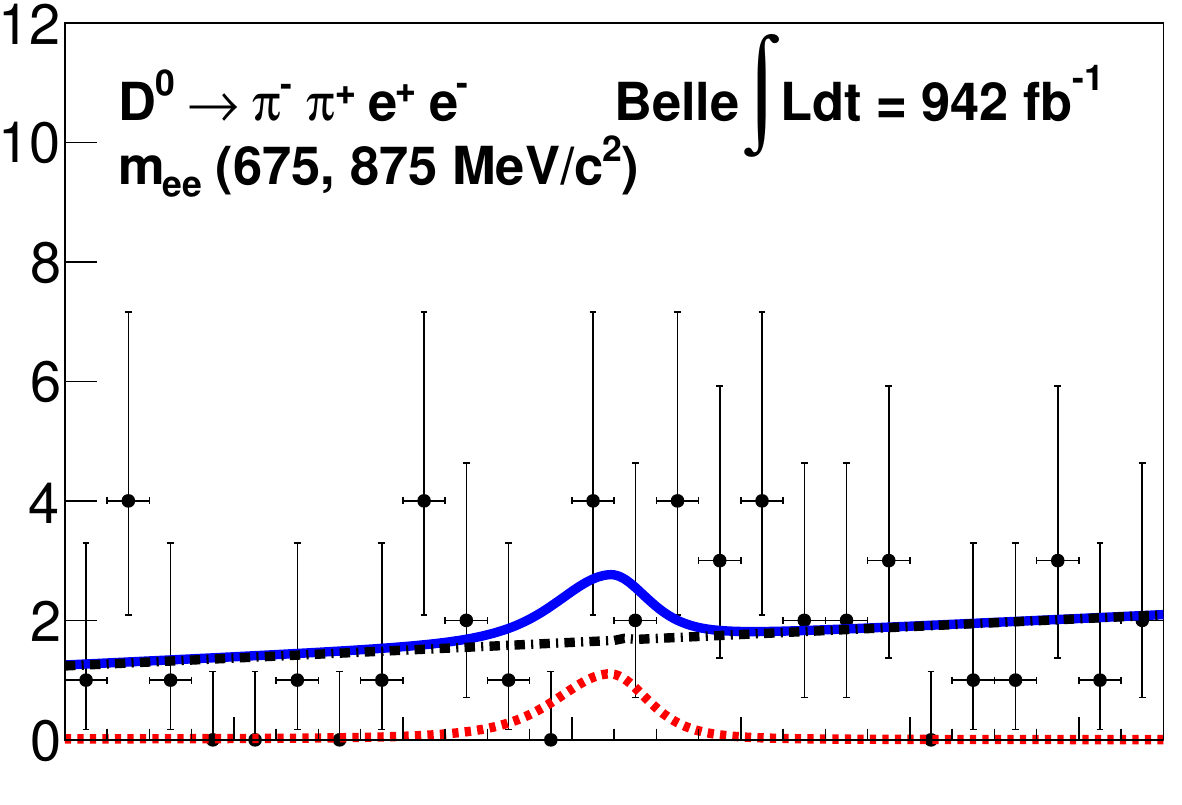}
      \end{minipage}
\hspace{-0.26cm} %
  \begin{minipage}{0.24\textwidth}
  \vspace{-0.19cm}
    \includegraphics[width=\textwidth]{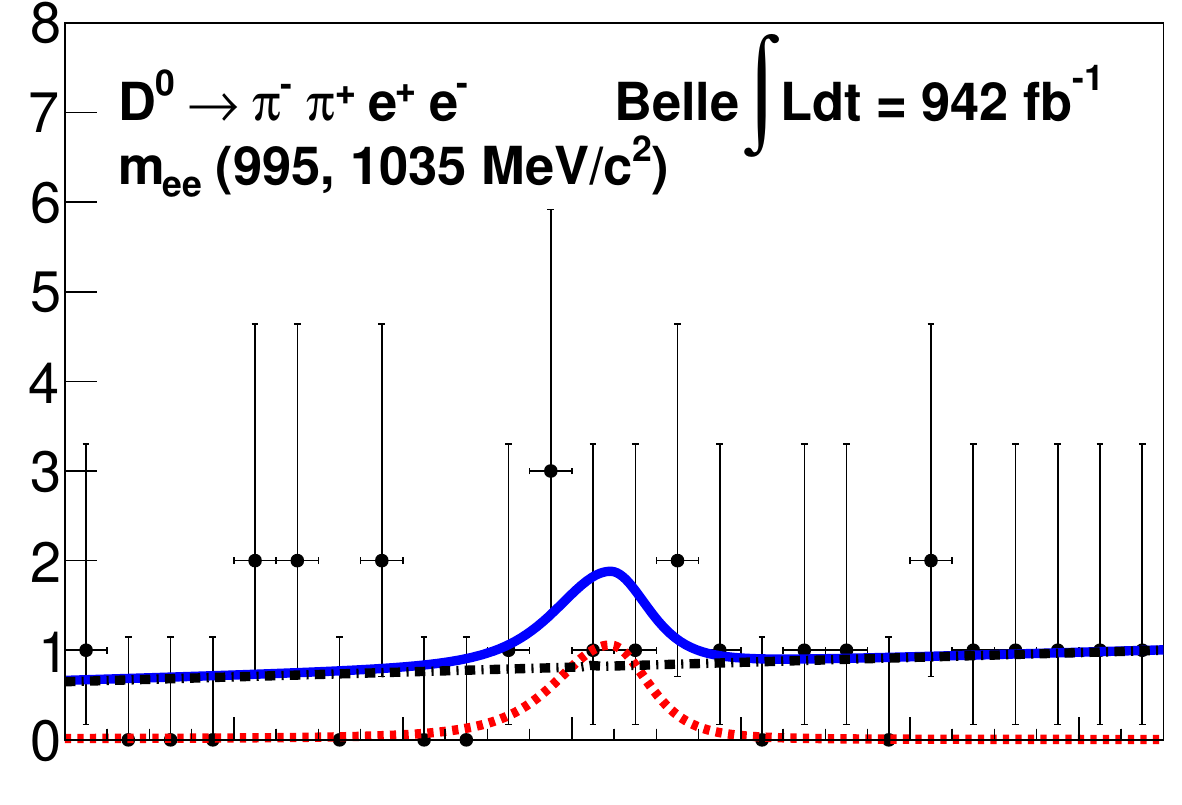}
  \end{minipage}
 \begin{minipage}{0.24\textwidth}
 \vspace{-0.19cm}
   {\includegraphics[width=\textwidth]{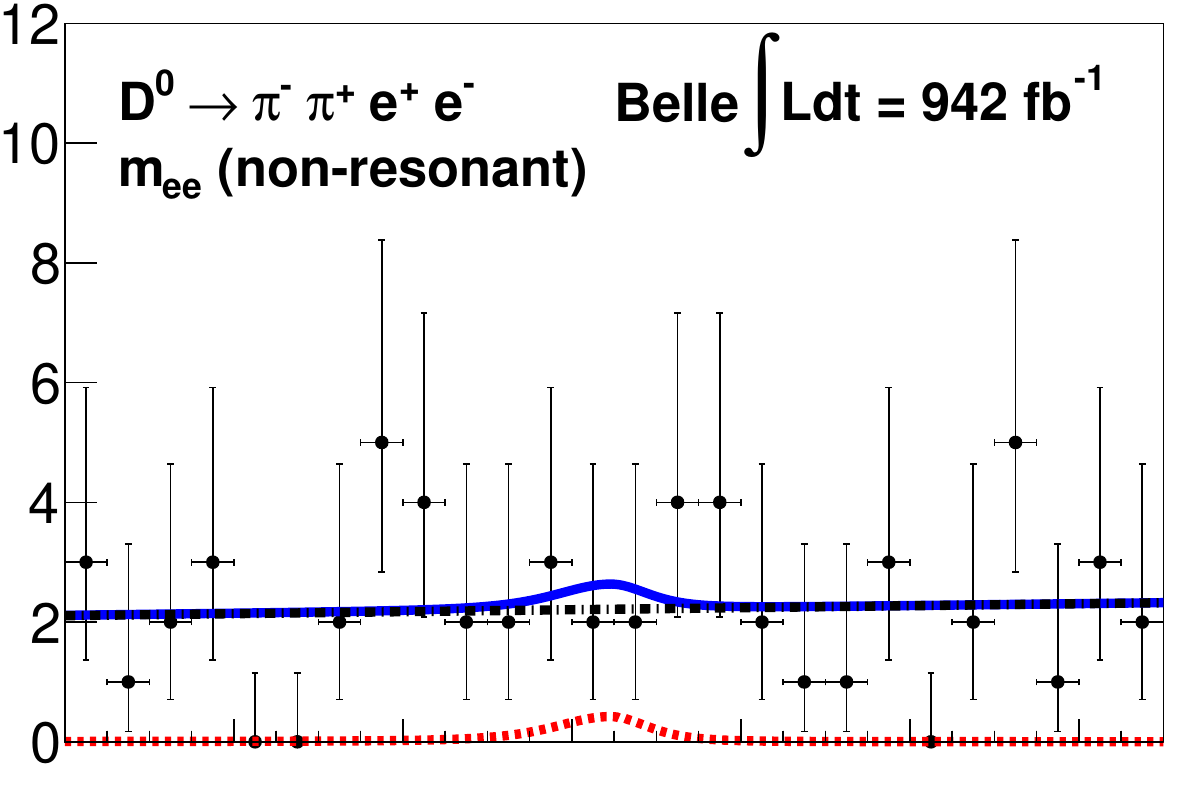}}
      \end{minipage}
\hspace{-0.26cm} %
  \begin{minipage}{0.24\textwidth}
  \vspace{-0.19cm} \includegraphics[width=\textwidth]{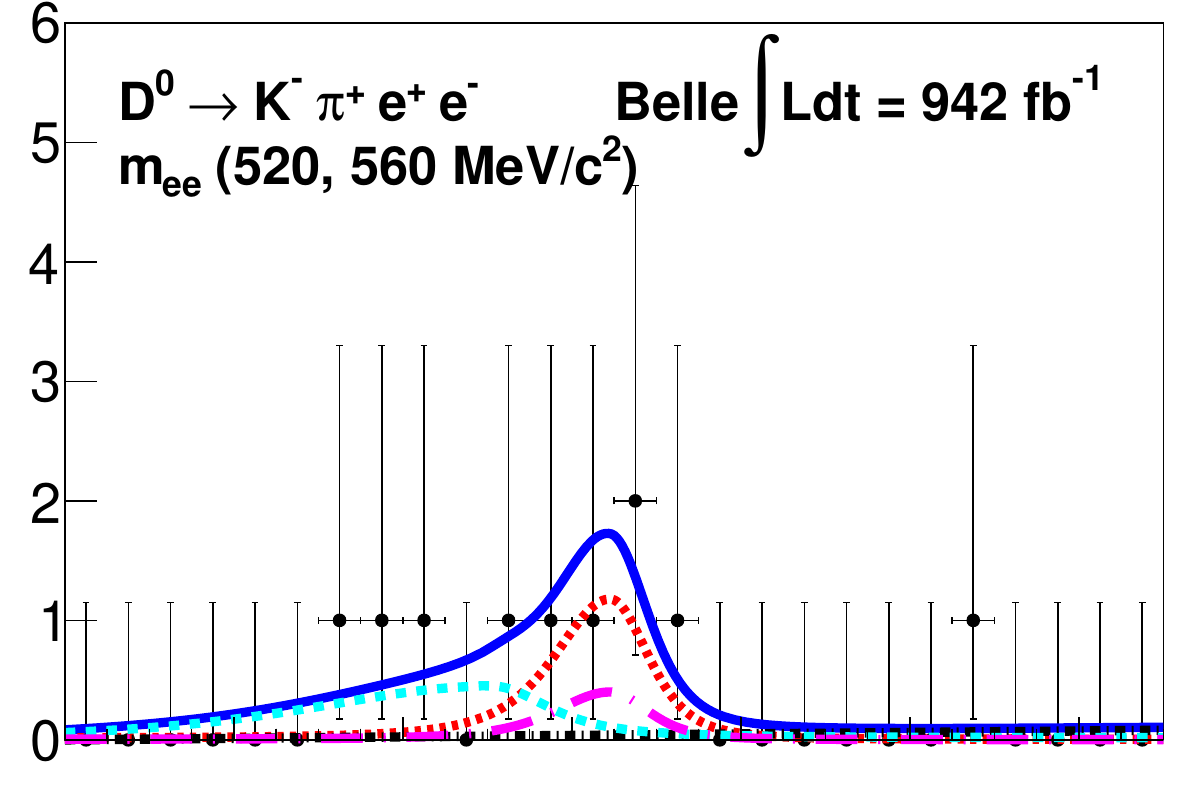}
  \end{minipage}
   \begin{minipage}{0.24\textwidth}
   \vspace{-0.17cm}
    \begin{overpic}[width=\linewidth]{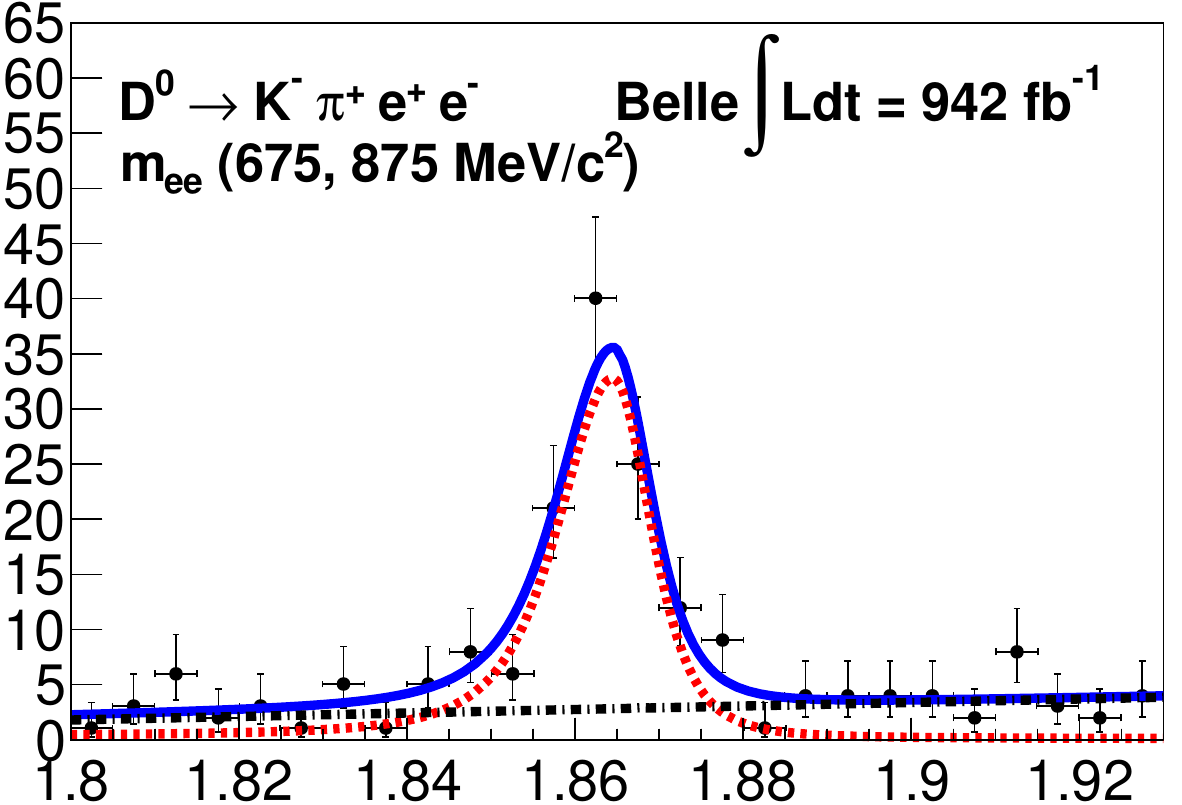}
    \put(46.6,-5.5){\text{\scriptsize{\mhhee $[\gevcc]$}}} 
    \end{overpic}
      \end{minipage}
      \hspace{-0.26cm}
  \begin{minipage}{0.24\textwidth}
  \vspace{-0.17cm}
   \begin{overpic}[width=\linewidth]{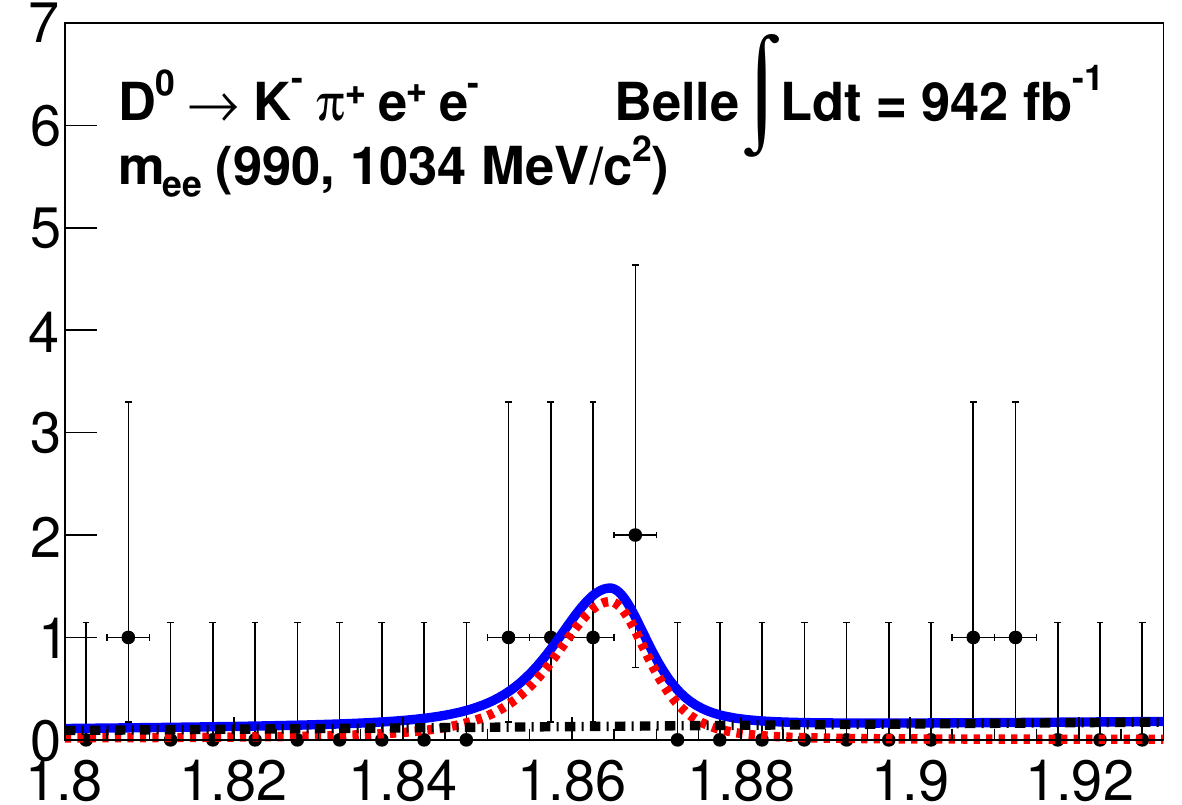}
  \put(46.6,-5.5){\text{\scriptsize{\mhhee $[\gevcc]$}}} 
    \end{overpic}
  \end{minipage}
  \vfill
  \hspace{-4.56cm} 
\begin{minipage}{0.24\textwidth}
\vspace{0.32cm}
 \includegraphics[width=\textwidth]{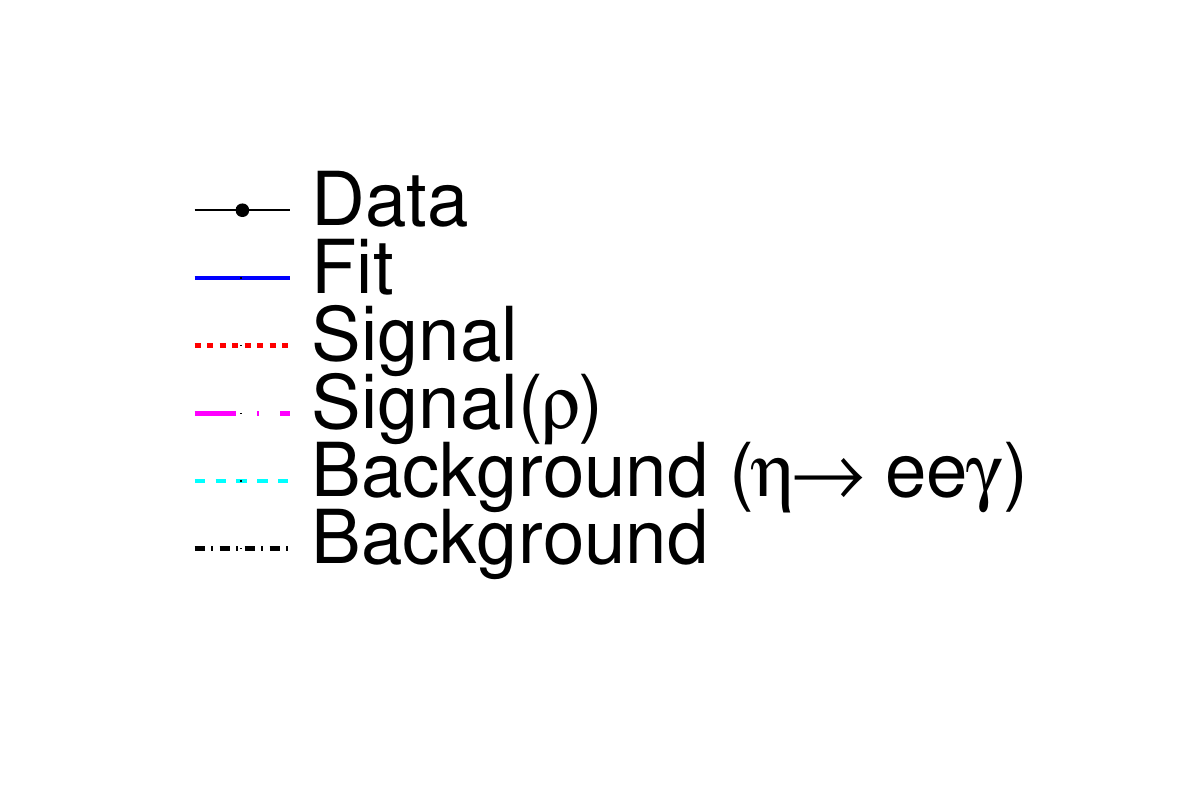}
\end{minipage}
\caption{\Dz\ra\hheered decays \mhhee distributions for \mee in the $\eta$, $\rho/\omega$ and $\phi$ mass regions. The $\eta\ra\epem\gamma$ decay background is shown for \Dz\ra\kpiee with \mee in the $\eta$ mass region (cyan dashed curve). These background PDF parameters are obtained from $\Dz\ra\Km\pip \eta\,\,[\ra\epem\gamma]$ MC simulation in which the $\gamma$ is not reconstructed. The \kkee mode with \mee in the $\eta$ mass region is not fitted since only one event is observed.} \label{fig-hheefit} 
\end{figure}

 We show the signal mode \mhhee distributions with projections of the fits superimposed for each \mee region in Fig.~\ref{fig-hheefit}. For each signal channel, we provide the branching fractions, and the corresponding significance $S = \sqrt{-2\Delta \mathrm{ln}\mathcal{L}}$, where $\Delta$ln$\mathcal{L}$ is the difference in the log-likelihood from the maximum value with respect to the value from the background-only hypothesis. We measure the branching fraction of \Dz\ra\kpiee in the \mee range $675 < \mee < 875$ \mevcc to be $(39.6 \pm 4.5 \pm 2.9)$ $\times$ 10$^{-7}$, where the first uncertainty is statistical and the second is systematic with a significance of 11.8$\sigma$. We set 90\% CL upper limits using the CL$_s$ method~\cite{ALRead:2002} for the channels with no significant signal; these results are in the range from $(2.3 - 8.1)$ $\times$ 10$^{-7}$. The extracted signal yields, significances, efficiencies, and branching fractions, or branching fraction upper limits for each \mee region are given in Table \ref{table-results}.\begin{table*}
\caption{\label{table-results}\Dz\ra\hhee yields, efficiencies, branching fractions, significances, and branching fraction upper limits @ 90\% CL of each \mee region. A fitted yield and a branching fraction are not reported for the \kkee mode with \mee in the $m_{\eta}$ region since only one event is observed, and the significance is determined from the CL$_s$ distribution. The first uncertainty is statistical and the second is systematic.} 
\begin{ruledtabular}
\begin{tabular}{ccccccc}
 \multicolumn{1}{c}{Decay mode} & \multicolumn{1}{c}{$\mee$ region $(\mevcc)$} & Yield & Efficiency (\%)  & \multicolumn{1}{c}{$\mathcal{B}$ $(\times 10^{-7}$)} & Significance & UL ($\times 10^{-7}$)\\ \hline
 \kkee  &\multicolumn{6}{c}{} \\ 
 $\eta$ & 520, 560 & \phantom{11}- & 3.33 $\pm$ 0.04 & - & $0.0\sigma$ & 2.3   \\
 $\rho^{0}/\omega$ & $>$ 675 \phantom{9}& \phantom{11}2.6 $\pm$\phantom{1}1.8 &  5.68 $\pm$ 0.06& \multicolumn{1}{c}{\phantom{1}1.2 $\pm$ 0.9 $\pm$ 0.1}  & 2.0$\sigma$ & 3.0   \\
 non-resonant & $> 200$ \footnote{Excluding resonance regions, which is the same for \pipiee modes.} & \phantom{11}3.5 $\pm$\phantom{1}3.3&  2.97 $\pm$ 0.04 & \multicolumn{1}{c}{\phantom{1}3.1 $\pm$ 3.0 $\pm$ 0.4} & 1.5$\sigma$ &  7.7  \\ \hline
 \pipiee  &\multicolumn{6}{c}{} \\ 
 $\eta$ & 520, 560 & \phantom{11}0.6 $\pm$\phantom{1}2.3 & 4.61 $\pm$ 0.05 &\multicolumn{1}{c}{\phantom{1}0.4 $\pm$ 1.4 $\pm$ 0.2}& 0.3$\sigma$ &  3.2   \\
 $\rho^{0}/\omega$ & 675, 875 & \phantom{11}3.7 $\pm$\phantom{1}4.1 &4.99 $\pm$ 0.05 & \multicolumn{1}{c}{\phantom{1}2.0 $\pm$ 2.2 $\pm$ 0.8} & 0.9$\sigma$ &  6.1   \\
 $\phi$ & \phantom{9}995, 1035 & \phantom{11}3.6 $\pm$\phantom{1}3.2 & 8.40 $\pm$ 0.06 & \multicolumn{1}{c}{\phantom{1}1.1 $\pm$ 1.1 $\pm$ 0.2} & 1.1$\sigma$ &  3.1   \\
 non-resonant & $> 200$\phantom{9} & \phantom{11}1.4 $\pm$\phantom{1}4.2 & 3.29 $\pm$ 0.04 & \multicolumn{1}{c}{\phantom{1}1.2 $\pm$ 3.4 $\pm$ 1.1} & 0.3$\sigma$ &  8.1   \\ \hline
 \kpiee  &\multicolumn{6}{c}{} \\ 
 $\eta$ & \multicolumn{1}{c}{520, 560} & \phantom{11}4.0 $\pm$\phantom{1}2.7 & 4.91 $\pm$ 0.04 & \multicolumn{1}{c}{\phantom{1}2.2 $\pm$ 1.5 $\pm$ 0.5}& 1.6$\sigma$ & 5.6   \\
 $\rho^{0}/\omega$ & 675, 875 & \phantom{11}110 $\pm$\phantom{1}13& 7.53 $\pm$ 0.06 &\multicolumn{1}{c}{39.6 $\pm$ 4.5 $\pm$ 2.9}& \phantom{1}11.8$\sigma$ & - \\
 $\phi$ & \phantom{9}990, 1034 & \phantom{11}4.6 $\pm$\phantom{1}2.4 & 8.75 $\pm$ 0.06&\multicolumn{1}{c}{\phantom{1}1.4 $\pm$ 0.8 $\pm$ 0.3}& 2.5$\sigma$ &  2.9   \\
\end{tabular}
\end{ruledtabular}
\end{table*} In the Supplemental Material, we show the projection of the fit on the \Dz\ra\hhee distribution as a function of $\mee^2$ with the background subtracted using the $s \mathcal{P}lot$ technique~\cite{PIVK2005356}. \\
\indent Systematic uncertainties can be divided into multiplicative and additive categories. The additive systematic uncertainties affect the determination of the signal and normalization mode yields and the corresponding significance. Multiplicative systematic uncertainties include PID and tracking efficiencies. The systematic uncertainty in tracking efficiency is 0.35\% per track, as determined from a control sample of $\Dstarp \ra \Dz(\KS\pip\pim) \pip$ decays. The systematic uncertainty due to \kaon identification is 1.0\%, determined from a study of a $\Dstarp \ra \Dz(\Km\pip) \pip$ control sample. The electron identification efficiency uncertainty is determined from the $\epem \ra \epem \epem$ processes and found to be 2.0\% for each track. The PID efficiency corrections are applied for the normalization mode and for each signal channel, and the particle identification systematics are about 5\%, which depend on the decay channel. We do not include a systematic uncertainty for the PID fake rates as the \Dz candidate invariant mass of misidentified \hhee decays do not peak near the \Dz mass after final selections according to MC simulations of hadronic \Dz decays. To account for the potential non-resonant decay contribution in the \mee resonance regions, the signal efficiency differences obtained using the signal MC between non-resonant and resonant decays are included in the systematic uncertainty. \\
\indent The uncertainty in yield extraction contributes to the additive systematic uncertainty, which affects the significance of the branching fraction. We obtain the PDF-related uncertainties by varying the PDF shapes and parameters for both signal and background. As alternative PDFs, we use two double-sided Crystal Ball functions~\cite{Skwarnicki:1986xj} with a shared mean for the signal and a second-order Chebyshev polynomial for the background PDF functions  to determine the signal yield systematics from the PDF shapes. In addition, the yield differences between the signal PDF parameters, fixed and floated, are incorporated into the systematic uncertainty. The additive systematic uncertainty for the background originating from the signal channel \Dz\ra\hhee with $ee$ from $\rho$ resonant decay is negligible for other $ee$ resonance regions. To incorporate the systematic uncertainties into the upper limits, the likelihood function is convolved with two Gaussian functions whose widths are the total multiplicative and additive systematic uncertainties and a third Gaussian with a width that is the sum in quadrature of the additive systematic uncertainties from the normalization mode. 
\par In summary, we have measured the branching fraction of \Dz\ra\kpiee in the \mee range $675 < \mee < 875$ \mevcc to be \begin{equation*}
    (39.6 \pm 4.5 \pm 2.9) \times 10^{-7}
\end{equation*} with a significance of 11.8$\sigma$ using 942 \invfb of Belle data. The measured branching fraction is consistent with and more precise than the \babar measurement~\cite{Lees:2018vns}. For the other $ee$ resonant and non-resonant regions, we do not observe any significant signal and set 90\% CL upper limits on the branching fractions. These limits range from 2.3 $\times$ 10$^{-7}$ to 8.1 $\times$ 10$^{-7}$. Our \Dz\ra\kpiee limits are more restrictive than the \babar~\cite{Lees:2019pej} and BES~III~\cite{Ablikim:2018gro} limits. \\
\\ \textit{Note added.}
\indent While this manuscript was being finalized, LHCb published new results on $\Dz\ra\pipiee$ and $\Dz\ra\kkee$ decays. They observe the former in two \mee mass regions and set upper limits on the latter that are 1.4--7.7 times more stringent than ours~\cite{LHCb2025}.\\
\section*{Acknowledgement}

This work, based on data collected using the Belle II detector, which was built and commissioned prior to March 2019,
and data collected using the Belle detector, which was operated until June 2010,
was supported by
Higher Education and Science Committee of the Republic of Armenia Grant No.~23LCG-1C011;
Australian Research Council and Research Grants
No.~DP200101792, 
No.~DP210101900, 
No.~DP210102831, 
No.~DE220100462, 
No.~LE210100098, 
and
No.~LE230100085; 
Austrian Federal Ministry of Education, Science and Research,
Austrian Science Fund (FWF) Grants
DOI:~10.55776/P34529,
DOI:~10.55776/J4731,
DOI:~10.55776/J4625,
DOI:~10.55776/M3153,
and
DOI:~10.55776/PAT1836324,
and
Horizon 2020 ERC Starting Grant No.~947006 ``InterLeptons'';
Natural Sciences and Engineering Research Council of Canada, Digital Research Alliance of Canada, and Canada Foundation for Innovation;
National Key R\&D Program of China under Contract No.~2024YFA1610503,
and
No.~2024YFA1610504
National Natural Science Foundation of China and Research Grants
No.~11575017,
No.~11761141009,
No.~11705209,
No.~11975076,
No.~12135005,
No.~12150004,
No.~12161141008,
No.~12405099,
No.~12475093,
and
No.~12175041,
and Shandong Provincial Natural Science Foundation Project~ZR2022JQ02;
the Czech Science Foundation Grant No. 22-18469S,  Regional funds of EU/MEYS: OPJAK
FORTE CZ.02.01.01/00/22\_008/0004632 
and
Charles University Grant Agency project No. 246122;
European Research Council, Seventh Framework PIEF-GA-2013-622527,
Horizon 2020 ERC-Advanced Grants No.~267104 and No.~884719,
Horizon 2020 ERC-Consolidator Grant No.~819127,
Horizon 2020 Marie Sklodowska-Curie Grant Agreement No.~700525 ``NIOBE''
and
No.~101026516,
and
Horizon 2020 Marie Sklodowska-Curie RISE project JENNIFER2 Grant Agreement No.~822070 (European grants);
L'Institut National de Physique Nucl\'{e}aire et de Physique des Particules (IN2P3) du CNRS
and
L'Agence Nationale de la Recherche (ANR) under Grant No.~ANR-23-CE31-0018 (France);
BMFTR, DFG, HGF, MPG, and AvH Foundation (Germany);
Department of Atomic Energy under Project Identification No.~RTI 4002,
Department of Science and Technology,
and
UPES SEED funding programs
No.~UPES/R\&D-SEED-INFRA/17052023/01 and
No.~UPES/R\&D-SOE/20062022/06 (India);
Israel Science Foundation Grant No.~2476/17,
U.S.-Israel Binational Science Foundation Grant No.~2016113, and
Israel Ministry of Science Grant No.~3-16543;
Istituto Nazionale di Fisica Nucleare and the Research Grants BELLE2,
and
the ICSC – Centro Nazionale di Ricerca in High Performance Computing, Big Data and Quantum Computing, funded by European Union – NextGenerationEU;
Japan Society for the Promotion of Science, Grant-in-Aid for Scientific Research Grants
No.~16H03968,
No.~16H03993,
No.~16H06492,
No.~16K05323,
No.~17H01133,
No.~17H05405,
No.~18K03621,
No.~18H03710,
No.~18H05226,
No.~19H00682, 
No.~20H05850,
No.~20H05858,
No.~22H00144,
No.~22K14056,
No.~22K21347,
No.~23H05433,
No.~26220706,
and
No.~26400255,
and
the Ministry of Education, Culture, Sports, Science, and Technology (MEXT) of Japan;  
National Research Foundation (NRF) of Korea Grants
No.~2021R1-F1A-1064008, 
No.~2022R1-A2C-1003993,
No.~2022R1-A2C-1092335,
No.~RS-2016-NR017151,
No.~RS-2018-NR031074,
No.~RS-2021-NR060129,
No.~RS-2023-00208693,
No.~RS-2024-00354342
and
No.~RS-2025-02219521,
Radiation Science Research Institute,
Foreign Large-Size Research Facility Application Supporting project,
the Global Science Experimental Data Hub Center, the Korea Institute of Science and
Technology Information (K25L2M2C3 ) 
and
KREONET/GLORIAD;
Universiti Malaya RU grant, Akademi Sains Malaysia, and Ministry of Education Malaysia;
Frontiers of Science Program Contracts
No.~FOINS-296,
No.~CB-221329,
No.~CB-236394,
No.~CB-254409,
and
No.~CB-180023, and SEP-CINVESTAV Research Grant No.~237 (Mexico);
the Polish Ministry of Science and Higher Education and the National Science Center;
the Ministry of Science and Higher Education of the Russian Federation
and
the HSE University Basic Research Program, Moscow;
University of Tabuk Research Grants
No.~S-0256-1438 and No.~S-0280-1439 (Saudi Arabia), and
Researchers Supporting Project number (RSPD2025R873), King Saud University, Riyadh,
Saudi Arabia;
Slovenian Research Agency and Research Grants
No.~J1-50010
and
No.~P1-0135;
Ikerbasque, Basque Foundation for Science,
State Agency for Research of the Spanish Ministry of Science and Innovation through Grant No. PID2022-136510NB-C33, Spain,
Agencia Estatal de Investigacion, Spain
Grant No.~RYC2020-029875-I
and
Generalitat Valenciana, Spain
Grant No.~CIDEGENT/2018/020;
the Swiss National Science Foundation;
The Knut and Alice Wallenberg Foundation (Sweden), Contracts No.~2021.0174, No.~2021.0299, and No.~2023.0315;
National Science and Technology Council,
and
Ministry of Education (Taiwan);
Thailand Center of Excellence in Physics;
TUBITAK ULAKBIM (Turkey);
National Research Foundation of Ukraine, Project No.~2020.02/0257,
and
Ministry of Education and Science of Ukraine;
the U.S. National Science Foundation and Research Grants
No.~PHY-1913789 
and
No.~PHY-2111604, 
and the U.S. Department of Energy and Research Awards
No.~DE-AC06-76RLO1830, 
No.~DE-SC0007983, 
No.~DE-SC0009824, 
No.~DE-SC0009973, 
No.~DE-SC0010007, 
No.~DE-SC0010073, 
No.~DE-SC0010118, 
No.~DE-SC0010504, 
No.~DE-SC0011784, 
No.~DE-SC0012704, 
No.~DE-SC0019230, 
No.~DE-SC0021274, 
No.~DE-SC0021616, 
No.~DE-SC0022350, 
No.~DE-SC0023470; 
and
the Vietnam Academy of Science and Technology (VAST) under Grants
No.~NVCC.05.02/25-25
and
No.~DL0000.05/26-27.

These acknowledgements are not to be interpreted as an endorsement of any statement made
by any of our institutes, funding agencies, governments, or their representatives.

We thank the SuperKEKB team for delivering high-luminosity collisions;
the KEK cryogenics group for the efficient operation of the detector solenoid magnet and IBBelle on site;
the KEK Computer Research Center for on-site computing support; the NII for SINET6 network support;
and the raw-data centers hosted by BNL, DESY, GridKa, IN2P3, INFN, 
PNNL/EMSL, 
and the University of Victoria.

\textit{Data availability.}\indent The full Belle II data are not publicly
available. The collaboration will consider requests for
access to the data that support this article.
\bibliography{references}
\clearpage
\onecolumngrid
\thispagestyle{empty}
\vspace{10pt}
\begin{center}
    \textbf{\large Supplemental material to $\Dz\ra\hhee$}
\end{center}
\vspace{10pt}
\setcounter{figure}{0}
\section*{\Dz\ra\hhee modes $dN/d\mee^{2}$ vs $\mee^{2}$ distributions}
Figure 1 (below) shows the \Dz\ra\hhee modes $dN/d\mee^{2} $ vs $\mee^{2}$ distributions with background subtracted using the $s\mathcal{P}lot$ technique.
\begin{figure}[!hbp]
  \begin{minipage}{0.495\textwidth}
    \includegraphics[width=\textwidth]{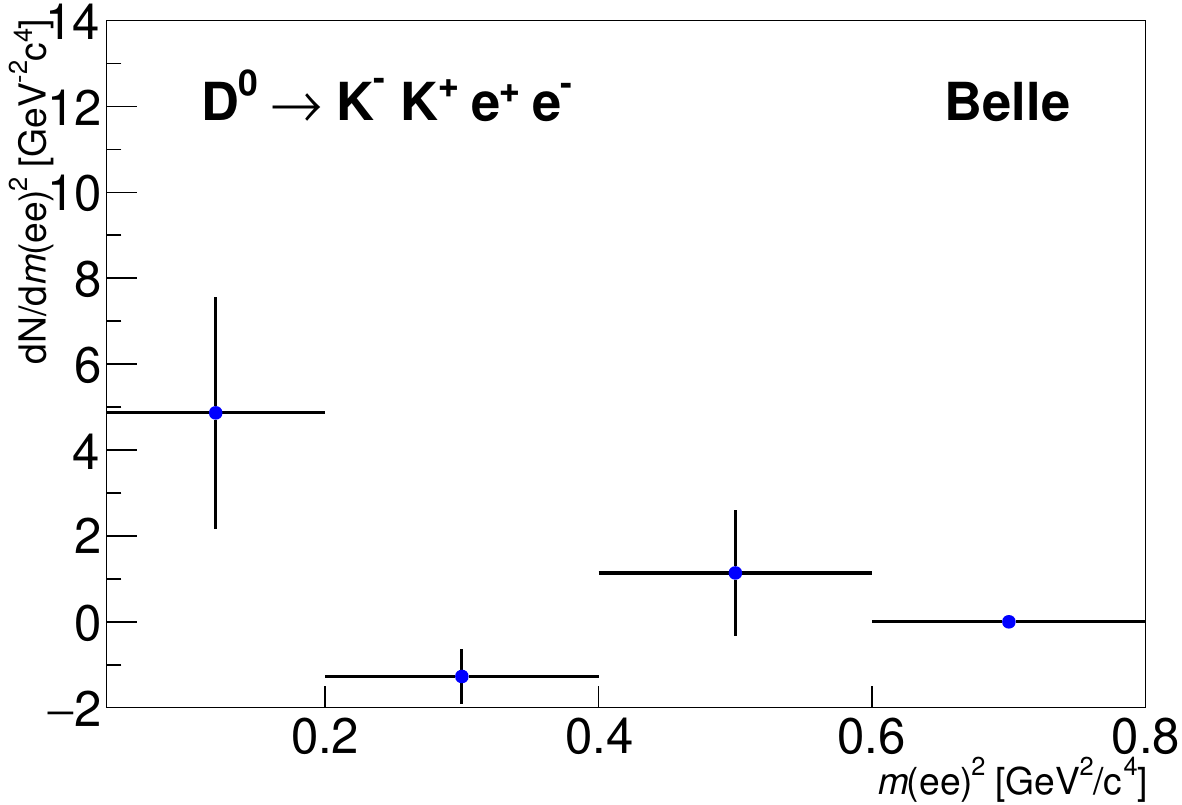}%
   \end{minipage}
\hspace{\fill} %
  \begin{minipage}{0.48\textwidth}
    \includegraphics[width=\textwidth]{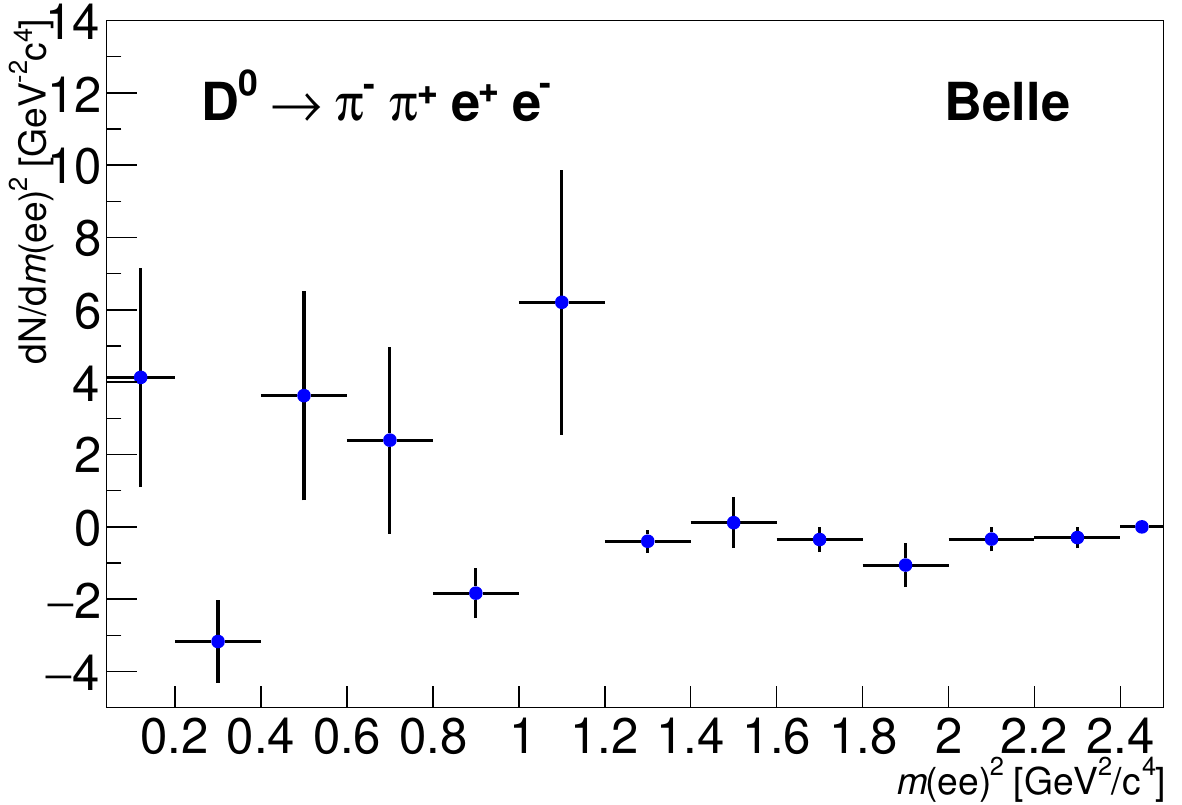}
  \end{minipage}
\hspace{\fill} %
  \begin{minipage}{0.5\textwidth}
      \includegraphics[width=\textwidth]{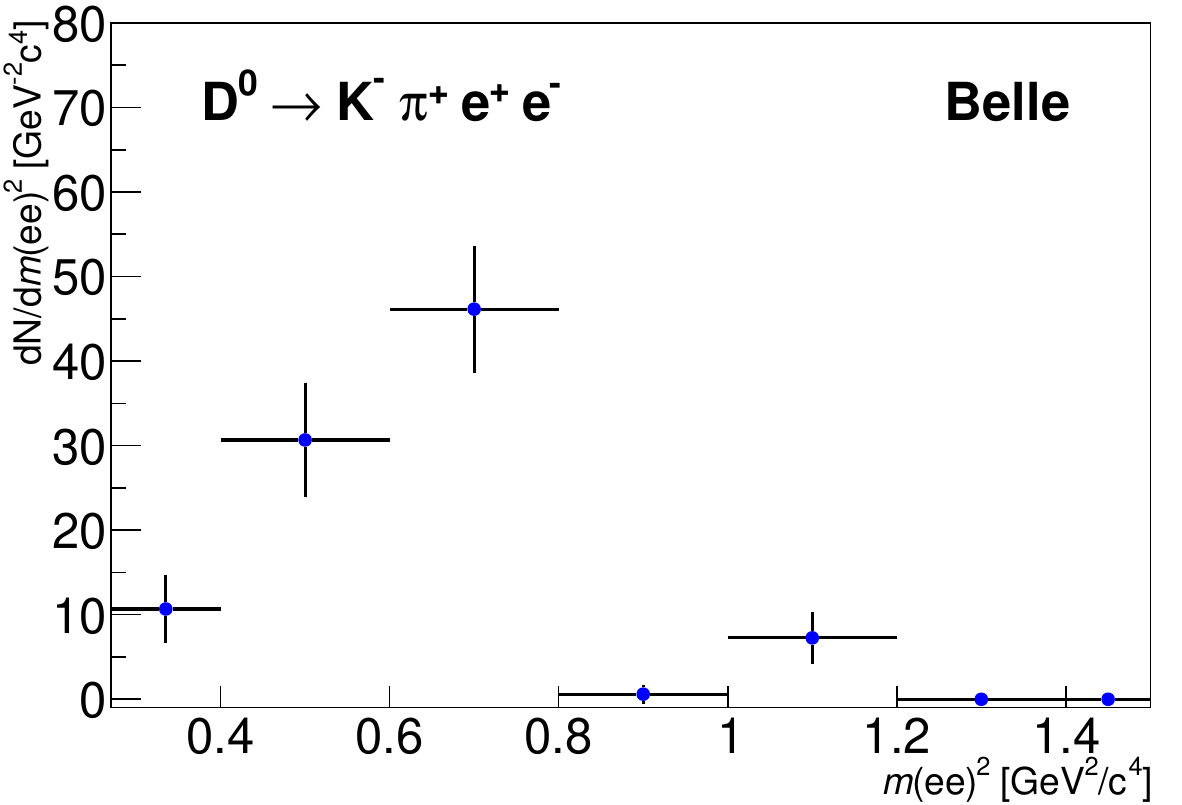}
  \end{minipage}
    
\caption{$dN/d\mee^{2} $ vs $\mee^{2}$ distributions for \Dz\ra\hhee candidates. The background has been subtracted using the $s\mathcal{P}lot$  technique. A significant \Dz\ra\kpiee signal with \mee in $\rho/\omega$ mass region is visible in the (0.4, 0.8\gevcccc) $\mee^{2}$ bins. The negative bins are due to low statistics after final selections. In the analysis, the selections are optimized separately in each resonance region and in the combined non-resonant regions. For these plots, the \Dscmp, \deltam, and PID selections for the non-resonant region are applied to all \mee regions. 
}
\end{figure}
\end{document}